# Collective Effects and Intense Beam-Plasma Interactions in Ion-Beam-Driven High Energy Density Matter and Inertial Fusion Energy

Igor D. Kaganovich (ikaganov@pppl.gov), Edward A. Startsev, Hong Qin, Erik Gilson

*Princeton Plasma Physics Laboratory, P.O. Box 451, Princeton NJ 08543*

Thomas Schenkel, Jean-Luc Vay, Ed P. Lee William Waldron, Roger Bangerter, Arun Persaud, Peter Seidl, and Qing Ji

*Accelerator Technology and Applied Physics Division, Lawrence Berkeley National Laboratory, Berkeley, CA 94720*

Alex Friedman, Dave P. Grote, John Barnard,

*Lawrence Livermore National Laboratory, P.O. Box 808, Livermore CA 94550*

**Executive summary**

For the successful generation of ion-beam-driven high energy density matter and heavy ion fusion energy, intense ion beams must be transported and focused onto a target with small spot size. One of the successful approaches to achieve this goal is to accelerate and transport intense ion charge bunches in an accelerator and then focus the charge bunches ballistically in a section of the accelerator that contains a neutralizing background plasma. This requires the ability to control space-charge effects during un-neutralized (non-neutral) beam transport in the accelerator and transport sections, and the ability to effectively neutralize the space charge and current by propagating the beam through background plasma. As the beam intensity and energy are increased in future heavy ion fusion (HIF) drivers and Fast Ignition (FI) approaches, it is expected that nonlinear processes and collective effects will become much more pronounced than in previous experiments. Making use of 3D electromagnetic particle-in-cell simulation (PIC) codes (BEST, WARP-X, and LTP-PIC, etc.), the theory and modelling studies will be validated by comparing with experimental data on the 100kV Princeton Advanced Test Stand, and future experiments at the FAIR facility. The theoretical predictions that are developed will be scaled to the beam and plasma parameters relevant to heavy ion fusion drivers and Fast Ignition scenarios. Therefore, the theoretical results will also contribute significantly toward the long-term goal of fusion energy production by ion-beam-driven inertial confinement fusion. The proposed research places special emphasis on addressing critical scientific issues in the following areas:

- Theoretically study collective beam-plasma interactions during longitudinal and transverse compression of the beam pulse for the HIF driver; and identify and mitigate the effects of collective beam-plasma interactions during compression of the beam pulse;
- Identify existing available ion beam facilities where modeling results can be validated;
- Develop, test and apply advanced plasma sources that produce sufficiently dense plasma for intense ion beam neutralization that are compatible with accelerator vacuum requirements;
- Experimentally validate the feasibility of heavy ion beams based on negative ions; develop reliable theoretical models describing the effects of ion-atom collisions on the lifetime of negative ions in the accelerator; and study the fundamental properties of ion-ion plasmas produced on the Princeton Advanced Test Stand (PATS) facility;
- Develop models describing the nonlinear dynamics of intense non-neutral ion beams with the goal of minimizing deleterious collective effects and instabilities and optimizing ion beam transport and focusing in the non-neutral section of the accelerator system;

- Develop and test innovative beam delivery and instability control techniques to maximize the focused beam intensity on target, including the use of oscillating wobbler fields in the final focus system for beam smoothing and facilitate uniform deposition on the target on FAIR facility, and scaling the concept to heavy ion fusion driver systems.

**Background and Motivation**

Charged particle beams produced in accelerators have a number of attractive features as potential drivers for IFE, and as a component in FI approaches [1,2]. To summarize the Refs. [1,2] HIF drivers can deliver high energy pulses more directly to a fusion target; they have high efficiency (>20% wall plug to beam); favorable final optic protection in a future reactor chamber; demonstrated long life of accelerator components; relatively high repetition rates (>10 Hz), etc. However, for HIF drivers the power requirements far exceed levels achieved in modern accelerators. White paper [2] is addressing ability to scale the accelerators to the HIF driver power levels.

A principal aim of the proposed research effort is to address the following compelling scientific questions as they pertain to ion-beam-driven High Energy Density Physics (HEDP) and HIF:

- How can heavy ion beams be compressed to the high intensities required to create high energy density matter and fusion conditions?
- How are intense charged-particle beams transported and focused?

An important long-term objective of the proposed U.S. HIF program is to provide a comprehensive scientific knowledge base and the enabling technologies required for IFE driven by high-brightness heavy ion beams, while the major near-term objective is to explore the limits of compressing ion charge bunches to very short pulses for purposes of investigating ion-driven HEDP and Warm Dense Matter (WDM) physics. A fundamental understanding of nonlinear space-charge effects on the propagation, acceleration and compression of high-brightness (high-current and low-emittance) heavy ion beams is essential to the identification of optimal operating regimes in which emittance growth and beam losses are minimized in periodic focusing accelerators and transport systems for applications of intense heavy ion beams to HEDP and IFE.

**Short Overview of the Current Status**

For the successful generation of ion-beam-driven high energy density matter and heavy ion fusion energy, intense ion beams must be transported and focused onto a target with small spot size. One of the successful approaches to achieve this goal is to accelerate and transport intense ion charge bunches in an accelerator and then focus the charge bunches ballistically in a section of the accelerator that contains neutralizing background plasma. This requires the ability to control space-charge effects during un-neutralized (non-neutral) beam transport in the accelerator and transport sections, and the ability to effectively neutralize the space charge and current by propagating the ion beam pulse through background plasma.

The High Current Experiment (HCX) was a very successful proof-of-principle experiment that demonstrated that intense ion beams with large space charge of up to several kV can be routinely transported and focused as un-neutralized beams to a spot size of a few mm diameter [3]. Another important series of experiments in HCX showed that electron clouds produced by background gas or surface ionization can be cleared from the ion beam pulse by making use of clearing electrodes or quadrupole magnetic fields, even for high-current ion beams [4].

The Neutralized Drift Compression Experiments -I and II (NDCX) were also a very successful proof-of-principle experiment that demonstrated the ability to longitudinally compress the beam current by a factor of up to 100, and transversely focus the neutralized ion beam from a 3 cm radius down to a few mm spot size [5]. By producing a background plasma with density large compared to the beam density, it was shown experimentally, theoretically, and in numerical simulations that the beam space charge and current can be neutralized to a sufficiently high degree to provide a very high degree of ballistic focusing [6].

The background plasma was created by ferroelectric plasma sources developed at PPPL which produced high density plasma (up to $10^{15}$ cm$^{-3}$) near the walls of the plasma source, and then filled the neutralized transport section of the NDCX-I device with plasma with density up to $10^{11}$ cm$^{-3}$ [7]. The plasma density in the central region was adequate for neutralized beam transport in the neutralized drift section of NDCX. There were four additional plasma sources based on vacuum cathodic arc technology that were employed near the target to increase the plasma density in the focal plane of the compressing beam pulse for better neutralization in this region. It was shown experimentally that individual plasma jets streaming from the periphery region of the strong solenoidal magnetic field near the focal plane can fill the center of the transport region with plasma and neutralize the ion beam pulse [8]. The mechanisms for plasma penetration across the magnetic field are still not fully understood. One possible mechanism is the time-dependent nature of the external solenoidal magnetic field, which is affected by eddy currents in the nearby metal walls. Another possible mechanism may be plasma instabilities that provide anomalous transport of the plasma across the magnetic field [9].

Theoretical analyses and PIC modeling have been performed to describe (i) beam transport in the unneutralized (nonneutral) section of the accelerator [10], and (ii) quasi-steady-state propagation of the neutralized beam pulse in a background plasma [11]. Possible collective instabilities associated with the large beam space charge in both the neutralized and unneutralized sections have been surveyed, and the linear growth of several collective instabilities have been calculated. For the most robust instabilities, the nonlinear stage of instability has been simulated using both particle-in-cell and nonlinear delta-f codes [12]. Examples include: the two-stream instability between the beam ions and the background plasma electrons in the neutralized drift section of the accelerator; and the Harris and Weibel instabilities in the unneutralized section of the accelerator. The Harris and Weibel instabilities can lead to longitudinal emittance growth due to a coupling through a collective 3D mode between the longitudinal and transverse degrees of freedom of particle motion. For the neutralized drift transport section, two-stream electrostatic and electromagnetic instabilities have also been analyzed [12]. For parameters in the NDCX and HCX experiments, it was shown that the growth rates (exponentiation lengths) were sufficiently slow (long) that these instabilities were not of concern in these experiments [13].

Experiments at PATS showed that very high degree of neutralization can be achieved due to accumulation of cold electrons in the beam path [14] and further confirmed by recent experiments [15].

Important details on limitation of neutralization of an ion beam pulse by emission by filaments was investigated recently making use of 2D [16] and 3D PIC [17]. It was found that solitary and surface waves are excited in the process of neutralization and can affect the remaining space charge. Another important finding was that 3D simulations can be needed to adequately describe the process [17]. It is timely that there has been robust development of high-performance PIC codes (Chaos [17], Warp-X [18], LTP-PIC [19]) that can be used effectively to simulate these complex phenomena. The white paper [20] describes planned modeling efforts using Warp-X code.

**Research Objectives**

As the beam intensity and energy in the HIF driver are increased compared to the previous NDCX experiments, it is expected that nonlinear processes and collective effects will become much more pronounced, therefore critical scientific issues needs to be addressed:

***High-brightness heavy ion beam transport in accelerator:*** Develop a basic understanding of the limits on beam space charge imposed by gas and electron cloud effects, together with beam matching and magnet nonlinearities, and determine the effects of collective interactions on beam quality and transport.

***Longitudinal and transverse compression of intense ion beams:*** Develop a basic understanding of the limits on longitudinal compression within neutralizing background plasma, and the effects of beam-plasma instabilities over distances required for focusing in the chamber. Develop a basic understanding of the limits on transverse compression and focal spot size set by chromatic aberrations due to uncompensated velocity spreads from upstream longitudinal compression, and the beam emittance growth from imperfect charge neutralization and beam-plasma interactions.

The theory and modeling studies need to be validated making use of experimental data on the 100kV Princeton Advanced Test Stand (PATS), past data available from the NDCX facility, and new experimental data to be obtained from experiments on the FAIR facility or others [1]. The theoretical predictions that are developed will be scaled to the beam and plasma parameters relevant to heavy ion fusion drivers. Therefore, the theoretical results will also contribute significantly toward the long-term goal of fusion energy production by ion-beam-driven ICF.

**Research Thrusts**

Emphasis is placed on the following Research Thrusts relevant to ion-beam-driven HEDP and HIF:

**Thrust Area #1** – Develop effective neutralization and focusing schemes for neutralized beam transport in HIF drivers

The key research objectives in this area are to design and test advanced plasma sources for robust neutralization of intense ion charge bunches and to perform a feasibility study of practical implementation of novel plasma sources for neutralized drift sections in current and future heavy ion accelerator systems with a long lifetime and low cost. Specific research tasks include: Demonstrate that the beam charge and current can be controlled or neutralized during neutralized drift compression by judicious choice of background plasma parameters; Demonstrate that collective instabilities can be controlled or mitigated during neutralized drift compression by profiling the plasma density or magnetic field; and Study the neutralization process and effects of transients on beam emittance during beam entry into the plasma.

A further research objective in this thrust area is to design and test advanced collective focusing schemes [21] in the PATS and other available high energy density laboratory physics facilities, which effectively utilize the large self-electric fields of the beam pulse and do not require large focusing magnets.

**Thrust Area #2** – Develop innovative beam driver concepts for energy delivery in heavy ion fusion systems using *negative ion beams* extracted from ion -- ion plasmas

In order to avoid possible electron cloud effects and charge exchange effects many linear accelerators use negative ions instead of positive ions. Consequently, the research objective in this thrust area is to develop a comprehensive justification for the use of negative ion beams for heavy ion fusion drivers and propose experimental tests of the concept based on halogen negative ions on the Princeton Advanced Test Stand [22].

**Thrust Area #3 –** Minimize deleterious collective effects and optimize ion beam transport and focusing in nonneutral section of accelerator system

The research objective in this thrust area is to demonstrate that the large space charge in the ion charge bunch can be transported quiescently through the non-neutral section of the accelerator system for driver-scale parameters. The specific research tasks include: Investigate transverse emittance growth due to coupling of collective modes with focusing element misalignments and beam mismatch; Study longitudinal emittance growth due in finite-length charge bunches due to collective effects and instabilities such as the Harris instability [23]; Design achromatic focusing system for simultaneous transverse and longitudinal compression on future heavy-ion-driver scale facility [24].

**Thrust Area #4 –** Develop innovative beam delivery and instability control techniques to optimize target performance

The research objective in this thrust area is to investigate the use of oscillating electric fields in wobbler system and innovative beam delivery methods for beam smoothing technique to mitigate instabilities and facilitate uniform deposition on the target. Specific research tasks include develop wobbler design concepts for beam smoothing and the Rayleigh-Taylor instability control in driver-scale systems [25].

**Conclusion**

Heavy-ion-beam drivers offer number of advantages as potential drivers for IFE. Initial experimental study successfully addressed critical issues of controlling instabilities and large space charge of the driver beam in past HCX and NDCX experiments. Further progress can be achieved by combined experimental, modeling and theoretical research efforts. Princeton Test Stand is available at PPPL and can be used to test innovative concepts such us collective focusing, negative ion beams, neutralization by filaments and undersense plasma [26], and for development innovative plasma sources for neutralization [27].

Future experiments on focusing of intense ion beams at FAIR [1], DARHT [28] and other facilities will provide sufficient data to validate well established theories of space charge and current neutralization by plasma and in combination with the use of high-performance PIC simulations [18, 19] will enable well-grounded designs of Heavy Ion Fusion Driver [2].

# References

[1] Thomas Schenkel, et *al.*, "Ion beams and Inertial Fusion Energy", White paper submitted to the IFE Science & Technology Community Strategic Planning Workshop https://lasers.llnl.gov/nif-workshops/ife-workshop-2022/white-papers (2022).

[2] R.O. Bangerter, et *al.,* "Systems Studies and Accelerator Design," White paper submitted to the IFE Science & Technology Community Strategic Planning Workshop https://lasers.llnl.gov/nif-workshops/ife-workshop-2022/white-papers (2022).

[3] P. A. Seidl, et *al.,* "The High Current Transport Experiment for Heavy-Ion Inertial Fusion," Proceedings of the 2003 Particle Accelerator Conference, pp. 536-540 (Portland, Oregon, May, 2003).

L. Prost, et al., "High current transport experiment for heavy ion inertial fusion", Physical Review Special Topics on Accelerators and Beams **8**, 020101 (2005).

[4] Michel Kireeff Covo, Arthur W. Molvik, Alex Friedman, Ronald Cohen, Jean-Luc Vay, Frank Bieniosek, David Baca, Peter A. Seidl, Grant Logan, Jasmina L. Vujic; "Electron cloud measurements in heavy-ion driver for HEDP and inertial fusion energy", Nuclear Instruments and Methods in Physics Research B **261**, 980–985 (2007).

[5] P. K. Roy, S. S. Yu, S. Eylon, et al., "Results on intense beam focusing and neutralization from the neutralized beam experiment," Physics of Plasmas **11**, 2890-2898 (2004).

[6] P. K. Roy, S. S. Yu, E. Henestroza, A. Anders, F. M. Bieniosek, J. Coleman, S. Eylon, W. G. Greenway, M. Leitner, B. G. Logan, W. L. Waldron, D. R. Welch, C. Thoma, A. B. Sefkow, E. P. Gilson, P. C. Efthimion, and R. C. Davidson, "Drift Compression of an Intense Neutralized Ion Beam," Physical Review Letters **95**, 234801 (2005).

W. L. Waldron, W. J. Abraham, D. Arbelaez, A. Friedman, J. E. Gavin, E. P. Gilson, W. G. Greenway, D. P. Grote, J.-Y. Jung, J. W. Kwan, M. Leitner, S. M. Lidia, T. M. Lipton, L. L. Reginato, M. J. Regis, P. K. Roy, W. M. Sharp, M. W. Stettler, J. H. Takakuwa, J. Volmering, V. K. Vytla, "The NDCX-II Engineering Design", Nuclear Instruments and Methods in Physics Reserach A **733**, 226 (2014).

P. A. Seidl, A. Persaud, W. L. Waldron, J. J. Barnard, R. C. Davidson, A. Friedman, E. P. Gilson, W. G. Greenway, D. P. Grote, I. D. Kaganovich, S. M. Lidia, M. Stettler, J. H. Takakuwa and T. Schenkel, "Short Intense Ion Pulses for Materials and Warm Dense Matter Research," Nuclear Instruments and Methods in Physics Research A **800**, 98 (2015).

P. A. Seidl, J. J. Barnard, R. C. Davidson, A. Friedman, E. P. Gilson, D. Grote, Q. Ji, I. D. Kaganovich, A. Persaud, W. L. Waldron and T. Schenkel, "Short-Pulse Compressed Ion Beams on the Neutralized Drift Compression Experiment (NDCX-II) Facility," J. Phys.: Conf. Ser. 717 012079 (2016).

P.A. Seidl, J.J. Barnard, E. Feinberg, A. Friedman, E.P. Gilson, D.P. Grote, Q. Ji, I.D. Kaganovich, B. Ludewigt, A. Persaud, C. Sierra, M. Silverman, A.D. Stepanov, A. Sulyman, F. Treffert, W.L. Waldron, M. Zimmer, and T. Schenkel, "Irradiation of materials with short, intense ion pulses at NDCX-II," Laser and Particle Beams **35**, 373-378 (2017).

[7] E. P. Gilson, R. C. Davidson, P. C. Efthimion, J. Z. Gleizer, I. D. Kaganovich, and Ya E. Krasik, "Plasma Source Development for the NDCX-I and NDCX-II Neutralized Drift Compression Experiments," Laser and Particle Beams **30**, 1 (2012).

[8] P. K. Roy, S. S. Yu, W. Waldron, A. Anders, D. Baca, F. M. Bieniosek, J. Coleman, R. C. Davidson, P. C. Efthimion, S. Eylon, E. P. Gilson, W. G. Greenway, E. Henestroza, I. Kaganovich, M. Leitner, B. G. Logan, A. B. Sefkow, P. Seidl, C. Thoma and D. R. Welch, "Neutralized Drift Compression Experiments (NDCX) with a High Intensity Ion Beam", Nuclear Instruments and Methods in Physics Research A **57**7, 223 (2007).

[9] G.D. Rossi, T.A. Carter, B. Seo, J. Robertson, M.J. Pueschel, and P.W. Terry, "Electromagnetic turbulence in increased β plasmas in the Large Plasma Device" Journal of Plasma Physics **87**, 905870401 (2021).

[10] A.V. Barkhudaryan, D.G. Koshkarev, A.N. Talyzin, "Studies on stability of charge-compensated ion beams", Fusion Engineering and Design **32-33**, 183-188 (1996).

M. Winkler, H. Wollnik, B. Pfreundtner, E.I. Escha, P. Spiller, "High current pulsed lenses for heavy ion fusion applications", Fusion Engineering and Design **32–33,** 385-389 (1996).

E. P. Lee, T. J. Fessenden, and L. J. Laslett, "Transportable Charge in a Periodic Alternating Gradient System," IEEE Transactions on Nuclear Science **NS-32**, 2489 (1998).

E. A. Startsev and R. C. Davidson, "Analytical Solutions for the Nonlinear Drift Compression (Expansion) of Intense Charged Particle Beams", New Journal of Physics **6**, 141 (2004).

E. P. Gilson, M. Chung, R. C. Davidson, P. C. Efthimion, R. Majeski and E. A. Startsev, "Simulation of Long-Distance Beam Propagation in the Paul Trap Simulator Experiment", Nuclear Instruments and Methods in Physics Research A**544**, 171 (2005).

H. Qin, R. C. Davidson, J.J. Barnard and E.P. Lee, "Drift Compression and Final Focus Options for Heavy Ion Fusion", Nuclear Instruments and Methods in Physics Research A**544**, 255 (2005).

S. M Lund and S. R. Chawla, "Space-charge transport limits of ion beams in periodic quadrupole focusing channels", Nuclear Instruments and Methods in Physics Research **A 561,** 203 (2006).

E. A. Startsev and S. M. Lund, "Approximate Analytical Solutions for Continuously Focused Beams and Single-Species Plasmas in Thermal Equilibrium", Phys. Plasmas **15**, 043101 (2008).

Moses Chung, Hong Qin, Ronald C. Davidson, "Envelope Hamiltonian for Charged-Particle Dynamics in General Linear Coupled Systems", Phys. Plasmas **23**, 074507 (2016).

M. Chung, H. Qin, R. C. Davidson, L. Groening, and C. Xiao, "Generalized Kapchinskij-Vladimirskij Distribution and Beam Matrix for Phase-Space Manipulations of High-Intensity Beams", Physical Review Letters **117**, 224801 (2016).

[11] I. D. Kaganovich, G. Shvets, E. Startsev and R. C. Davidson, "Nonlinear Charge and Current Neutralization of an Intense Ion Beam Pulse in a Preformed Plasma," Phys. Plasmas **8**, 4180 (2001).

D. V. Rose, D. R. Welch, B. V. Oliver, R. E. Clark, W. M. Sharp, and A. Friedman, "Ballistic-neutralized chamber transport of intense heavy ion beams", Nuclear Instruments and Methods in Physics Research Section A: Accelerators, Spectrometers, Detectors and Associated Equipment, **464**, 299–304 (2001).

W. M. Sharp, D. A. Callahan, M. Tabak, S. S. Yu, and P. F. Peterson, "Chamber transport of "foot" pulses for heavy-ion fusion", Physics of Plasmas **10**, 2457 (2003).

W. M. Sharp, D. A. Callahan, M. Tabak, S. S. Yu, P. F. Peterson, D. V. Rose, and D. R. Welch, "Chamber-transport simulation results for heavy-ion fusion drivers", Nuclear Fusion **44**, S221, 11 (2004).

I.D. Kaganovich, E. A. Startsev, R. C. Davidson and D. R. Welch, "Ion Beam Pulse Neutralization by a Background Plasma in a Solenoidal Magnetic Field," Nuclear Instruments and Methods in Physics Research A **544**, 383 (2005).

I. D. Kaganovich, A. B. Sefkow, E.A. Startsev, R. C. Davidson and D. R. Welch, "Effects of Finite Pulse Length, Electron Temperature, Magnetic Field, and Gas Ionization on Ion Beam Pulse Neutralization by Background Plasma", Nuclear Instruments and Methods in Physics Research A **577**, 93 (2007).

A. B. Sefkow, R. C. Davidson, E. P. Gilson, I. D. Kaganovich, A. Anders, J. E. Coleman, M. Leitner, S. M. Lidia, P. K. Roy, P. A. Seidl, W. L. Waldron, S. S. Yu, and D. R. Welch, "Simulations and experiments of intense ion beam current density compression in space and time," Physics of Plasmas **16**, 056701 (2009).

I. D. Kaganovich, R. C. Davidson, M. A. Dorf, E. A. Startsev, A. B Sefkow, A. F. Friedman and E. P. Lee, "Physics of Neutralization of Intense High-Energy Ion Beam Pulses by Electrons", Physics of Plasmas **17**, 056703 (2010).

M. Dorf, I. Kaganovich, E. Startsev, and R. C. Davidson, "Whistler Wave Excitation and Effects of Self-Focusing on Ion Beam Propagation through a Background Plasma along a Solenoidal Magnetic Field", Physics of Plasmas **17**, 023103 (2010).

[12] E. A. Startsev and R. C. Davidson, "Electromagnetic Weibel Instability in Intense Charged Particle Beams With Large Temperature Anisotropy", Phys. Plasmas **10**, 4829 (2003).

H. S. Uhm and R. C Davidson, "Effects of Electron Collisions on the Resistive Hose Instability in Intense Charged Particle Beams Propagating Through Background Plasma", Physical Review Special Topics on Accelerators and Beams **6**, 034204 (2003).

R. C. Davidson, I Kaganovich, H. Qin, E. A. Startsev, D. R. Welch, D. V. Rose and H. S. Uhm "Survey of Collective Instabilities and Beam-Plasma Interactions for Heavy Ion Beams", Physical Review Special Topics on Accelerators and Beams **7**, 114801 (2004).

E.A. Startsev, R.C. Davidson and H. Qin, "Anisotropy-Driven Collective Instability in Intense Charged Particle Beams", Physical Review Special Topics in Accelerators and Beams **8**, 124201 (2005).

H.S. Uhm and R. C. Davidson, "Theory of Resistive Hose Instability in Intense Charged Particle Beams Propagating Through Background Plasma with Low Electron Collision Frequency", IEEE Transactions in Plasma Science **33**, 1395 (2005).

E. A. Startsev and R. C. Davidson, "Two-Stream Instability for a Longitudinally-Compressing Charged Particle Beam", Physics of Plasmas **13**, 062108 (2006).

R. C. Davidson, M. Dorf, I. D. Kaganovich, H. Qin, A. B. Sefkow and Startsev, D. R. Welch, D. D. Rose and S. M. Lund, "Multspecies Weibel Instability for Intense Charged Particle Beam Propagation Through Background Plasma", Nuclear Instruments and Methods in Physics Research A**577**, 70 (2007).

H. Qin, R. C. Davidson and E. A. Startsev, "Nonlinear Delta-f Particle Simulations of Collective Effects in High-Intensity Bunched Beams", Nuclear Instruments and Methods in Physics Research **A 577,** 86 (2007).

E. A. Startsev and R. C. Davidson, "Dynamic Stabilization of the Two-Stream Instability During Longitudinal Compression of Intense Charged Particle Beam Propagation Through Background Plasma", Nuclear Instruments and Methods in Physics Research A**577**, 79 (2007).

G. Shvets, O. Polomarov, V. Khudik, C. Siemon and I. Kaganovich, "Nonlinear Evolution of the Weibel Instability in Relativistic Electron Beams", Phys. Plasmas **16**, 056701 (2009).

R. C. Davidson, M. A. Dorf, I. D. Kaganovich, H. Qin, A. B. Sefkow, E. A. Startsev, D. R. Welch, D. V. Rose, and S. M. Lund "Survey of Collective Instabilities and Beam-Plasma Interactions in Intense Heavy Ion Beams", Nuclear Instruments and Methods in Physics Research **A 606,** 11 (2009).

Ruili Zhang, Hong Qin, Ronald C. Davidson, Jian Liu, Jianyuan Xiao, "On the structure of the two-stream instability -- complex G-Hamiltonian structure and Krein collisions between positive- and negative-action modes", Physics of Plasmas **23**, 072111 (2016).

Kentaro Hara, Igor D Kaganovich, and Edward A Startsev, "Generation of forerunner electron beam during interaction of ion beam pulse with plasma", Phys. Plasmas **25**, 11609 (2018).

[13] E. A. Startsev, R. C. Davidson and M. Dorf, "Streaming Instabilities of Intense Charged Particle Beams Propagating along a Solenoidal Magnetic Field in a Background Plasma", Physics of Plasmas **15**, 062107 (2008).

[14] A.D. Stepanov, E.P. Gilson, L.R. Grisham, I.D. Kaganovich, and R.C. Davidson, "Dynamics of ion beam charge neutralization by ferroelectric plasma sources," Physics of Plasmas **23**, 122111 (2016).

A.D. Stepanov, J.J. Barnard, A. Friedman, E.P. Gilson, D.P. Grote, Q. Ji, I.D. Kaganovich, A. Persaud, P.A. Seidl, and T. Schenkel, "Optimizing beam transport in rapidly compressing beams on the neutralized drift compression experiment-II," Matter and Radiation at Extremes **3**, 78 (2018).

[15] A.V.Stepanov, Haowen Zhong, Zhang Shijian, Mofei Xub Xiaoyun Leb, G.E.Remnev, "Study of the propagation of an intense ion beam to the target", Vacuum **198**, 110892 (2022).

[16] C. Lan, and I.D. Kaganovich, "Electrostatic solitary waves in ion beam neutralization," Physics of Plasmas **26**, 050704 (2019).

C. Lan, and I.D. Kaganovich, "Neutralization of ion beam by electron injection: excitation and propagation of electrostatic solitary waves," Physics of Plasmas **27**, 043104 (2020).

C. Lan, and I.D. Kaganovich, "Neutralization of ion beam by electron injection: accumulation of cold electrons" Physics of Plasmas **27**, 043108 (2020).

[17] Nakul Nuwal, Deborah A. Levin, and Igor D. Kaganovich, "Excitation of three-dimensional electrostatic solitary waves and surface waves in beam neutralization," to be submitted to Phys. Rev. E (2022).

[18] WarpX Source and Documentation: https://ecp-warpx.github.io

J.-L. Vay, A. Almgren, J. Bell, L. Ge, D. P. Grote, M. Hogan, O. Kononenko, R. Lehe, A. Myers, C.Ng, J. Park, R. Ryne, O. Shapoval, M. Thévenet, W. Zhang "Warp-X: A new exascale computing platform for beam–plasma simulations", Nucl. Inst. Meth. A **909**, 486-479 (2018)

A. Myers, A. Almgren, L. D. Amorim, J. Bell, L. Fedeli, L. Ge, K. Gott, D. P. Grote, M. Hogan, A. Huebl, R. Jambunathan, R. Lehe, C. Ng, M. Rowan, O. Shapoval, M. Thévenet, J.-L. Vay, H. Vincenti, E. Yang, N. Zaim, W. Zhang, Y. Zhao and E. Zoni, "Porting WarpX to GPU-accelerated platforms", Parallel Computing **108**, 102833 (2021)

J.-L. Vay, A. Huebl, A. Almgren, L. D. Amorim, J. Bell, L. Fedeli, L. Ge, K. Gott, D. P. Grote, M. Hogan, R. Jambunathan, R. Lehe, A. Myers, C. Ng, M. Rowan, O. Shapoval, M. Thevenet, H. Vincenti, E. Yang, N. Zaïm, W. Zhang, Y. Zhao, and E. Zoni, "Modeling of a chain of three plasma accelerator stages with the WarpX electromagnetic PIC code on GPUs", Phys. Plasmas **28**, 023105 (2021) Special Collection: Building the Bridge to Exascale Computing: Applications and Opportunities for Plasma Science.

[19] LTP-PIC is PIC code on hybrid GPU-CPU architecture built using Open ACC; it is under development but it already participated in international 2D study where it showed superior performance, T. Charoy, J. Boeuf, A. Bourdon, J.A. Carlsson, P. Chabert, B. Cuenot, D. Eremin, L. Garrigues, K. Hara, I.D. Kaganovich, A.T. Powis, A. Smolyakov, D. Sydorenko, A. Tavant, O. Vermorel, and W. Villafana, "2D axial-azimuthal particle-in-cell benchmark for low-temperature partially magnetized plasmas," Plasma Sources Science and Technology **28**, 105010 (2019).

[20] J.-L. Vay, A. Friedman, E. P. Lee, P. A. Seidl, and J. J. Barnard., "Propagation of Ion Beams in a Heavy-Ion Inertial Fusion System," White papers submitted to the IFE Science & Technology Community Strategic Planning Workshop https://lasers.llnl.gov/nif-workshops/ife-workshop-2022/white-papers.

[21] M. Dorf, I.D. Kaganovich, E.A. Startsev, and R.C. Davidson, "Collective focusing of intense ion beam pulses for high-energy density physics applications," Physics of Plasmas **18**, 033106 (2011).

M.A. Dorf, R.C. Davidson, I.D. Kaganovich, and E.A. Startsev, "Enhanced collective focusing of intense neutralized ion beam pulses in the presence of weak solenoidal magnetic fields," Physics of Plasmas **19**, 056704 (2012).

[22] L. R. Grisham, E. P. Gilson, I. Kaganovich, J. W. Kwan and A. Stepanov, "Experimental Program for the Princeton Ion Source Test Facility", Laser and Particle Beams **10**, 571 (2010).

[23] Edward A. Startsev, Ronald C. Davidson, and Hong Qin, "Collective temperature anisotropy instabilities in intense charged particle beams" Physics of Plasmas **14**, 056705 (2007).

E. A. Startsev, R. C. Davidson and M. Dorf, "Approximate Matched Kinetic Quasi-Equilibrium Solutions for an Intense Charged Particle Beam Propagating Through a Periodic Focusing

Quadrupole Lattice", Physical Review Special Topic on Accelerators and Beams **13**, 064402 (2010).

M. A. Dorf, R. C. Davidson and E. A. Startsev, A Spectral Method for Halo Particle Definition in Intense Mismatched beams" Physics of Plasmas **18**, 043109 (2011).

E. A. Startsev, R. C. Davidson and M. Dorf, "Novel Hamiltonian Method for Collective Dynamics Analysis of an Intense Charged Particle Beam Propagating Through a Periodic Focusing Quadrupole Lattice," Physics of Plasmas **18**, 056712 (2011).

[24] H. Qin, R. C. Davidson, J. J. Barnard and E. P. Lee, "Drift Compression and Final Focus for Intense Heavy Ion Beams with Non-periodic, Time-Dependent Lattice", Physical Review Special Topics on Accelerators and Beams **7**, 104201 (2004).

J. M. Mitrani, I. D. Kaganovich and R. C. Davidson, "Mitigating Chromatic Effects on the Transverse Focusing of Intense Charged Particle Beams for Heavy Ion Fusion," Nuclear Instruments and Methods in Physics Research A **733**, 65 (2014)

[25] H. Qin, R. C. Davidson and B. G. Logan, "Centroid and Envelope Dynamics of High-Intensity Charged Particle Beams in an External Focusing Lattice and Oscillating Wobbler", Physical Review Letters **104**, 254801 (2010).

H. Qin, R. C. Davidson and B. G. Logan, Centroid and Envelope Dynamics of High-Intensity Charged Particle Beams in an External Focusing Lattice and Oscillating Wobbler for Heavy Ion Fusion", Laser and Particle Beams **29**, 365 (2011).

[26] W. Berdanier, P.K. Roy, and I.D. Kaganovich, "Intense ion beam neutralization using underdense background plasma," Physics of Plasmas **22**, 013104 (2015).

[27] E. P. Gilson, R. C. Davidson, P. C. Efthimion, I. D. Kaganovich, J. Gleizer, Y. Krasik, "Plasma source development for the NDCX-I and NDCX-II neutralized drift compression experiments", Laser and Particle Beams **30**, 435-443 (2012).

E. P. Gilson, R. C. Davidson, P. C. Efthimion, I. D. Kaganovich, J. W. Kwan, S. M. Lidia, P. A. Ni, P. K. Roy, P. A. Seidl, W. L. Waldron, J. J. Barnard and A. Friedman, "Ferroelectric Plasma Sources for NDCX-II and Heavy Ion Drivers," Nuclear Instruments and Methods in Physics Research A **733**, 226 (2014).

L. R. Grisham, A. Von Halle, A. F. Carpe, Guy Rossi, K. R. Gilton, E. D. McBride, E. P. Gilson, A. Stepanov, and T. N. Stevenson, "Studies of Electrical Breakdown Processes across Vacuum Gaps," Nuclear Instruments and Methods in Physics Research, A **733**, 168 (2014).

J. Z. Gleizer, Ya. E. Krasik, U. Dai and J.G. Leopold, "Vacuum surface flashover: experiments and simulations", Trans. On Dielectrics and Electrical Insulation **21**, 2394 (2014).

[28] https://www.lanl.gov/science-innovation/science-facilities/DARHT/